%% file: main.tex
\pdfoutput=1
\documentclass[conference]{IEEEtran}
\IEEEoverridecommandlockouts

\usepackage{cite}
\usepackage{amsmath,amssymb,amsfonts}
\usepackage[noend]{algorithmic}

\usepackage{algorithm}
\usepackage{graphicx}
\usepackage{tikz}
\usetikzlibrary{positioning}
\usepackage{booktabs}
\usepackage{xcolor}
\usepackage[hyphens]{url}
\usepackage[hidelinks]{hyperref}

\graphicspath{{1-intro/figs/}{3-problem/figs/}{4-racorn/figs/}{5-deadline/figs/}{6-eval/figs/}}

\newcommand{\Mzero}{M_{0}}
\newcommand{\Mavg}{M_{\mathrm{avg}}}
\newcommand{\ef}{\mathit{ef}}
\newcommand{\br}{b}
\newcommand{\sel}{s}
\newcommand{\racorn}{RACORN}
\DeclareRobustCommand{\racornone}{\ifmmode RACORN-1\else RACORN\nobreakdash-1\fi}
\DeclareRobustCommand{\acornone}{\ifmmode ACORN-1\else ACORN\nobreakdash-1\fi}

\begin{document}

\bstctlcite{IEEEexample:BSTcontrol}

\title{RACORN: Robust Predicate-Agnostic Filtered Vector Search under Extreme Selectivity}

\author{%
\IEEEauthorblockN{Yoonseok Kim}
\IEEEauthorblockA{\textit{Naver Corporation} \\
Seongnam, South Korea \\
kys910524@gmail.com \\
ORCID 0009-0003-1219-2089}
\and
\IEEEauthorblockN{Gyusik Choe}
\IEEEauthorblockA{\textit{Naver Corporation} \\
Seongnam, South Korea \\
kyusic@gmail.com \\
ORCID 0009-0002-4356-4079}}

\maketitle

\begin{abstract}
\input{0-abstract/abstract}
\end{abstract}


\section{Introduction}\label{sec:intro}
\input{1-intro/intro}

\section{Background}\label{sec:background}
\input{2-background/background}

\section{Problem Analysis}\label{sec:problem}
\input{3-problem/problem}

\section{RACORN: Adaptive Search Fallback}\label{sec:racorn}
\input{4-racorn/racorn}
\subsection{Cost and Recall, Cycle by Cycle}\label{sec:cost}
\input{4-racorn/cost}

\section{Search Under a Deadline}\label{sec:deadline}
\input{5-deadline/deadline}

\section{Evaluation}\label{sec:eval}
\input{6-eval/eval}

\section{Conclusion}\label{sec:conclusion}
\input{8-conclusion/conclusion}

\section*{Acknowledgment}
\input{9-ack/ack}

\bibliographystyle{IEEEtran}
\bibliography{refs}

\end{document}

%% file: 0-abstract/abstract.tex
Filtered vector search returns the nearest neighbors among the vectors that
pass a predicate. The predicate is often selective: most tag queries
of the Big ANN filtered track select less than 1\% of YFCC~10M. Predicate-agnostic
methods search one HNSW graph built without metadata and so take any
predicate, but below a few percent selectivity they break down. \acornone{}
draws candidates only from passing neighbors and its recall collapses below
1\%. HNSW keeps searching but spends about $\ef/\sel$ expansion cycles at
selectivity $\sel$. We present \racornone{}, an extension of \acornone{} that
keeps its graph and filter-first traversal and adds Adaptive Search Fallback.
When the two-hop expansion of a node falls short of a target, failing nodes
enter the candidate queue as bridges, under a per-expansion budget, but never
enter the result queue. \racornone{} serves the selectivities \acornone{} cannot:
from 1\% down \acornone{} no longer reaches recall 0.9 at any $\ef$, while
\racornone{} reaches 0.95 from 3\% down on every dataset we sweep. It also reaches 0.9 on every natural filter that YFCC
and LAION ship and on every conjunction and caption predicate we run, where
\acornone{} misses more than half. With a deadline equal to the time HNSW
needs to fill its result queue, \racornone{} returns the higher recall on all
eight datasets at four selectivities. On eight indexes of 1M to 100M vectors, at
1\% selectivity and recall 0.9 it is 1.3--8.4$\times$ faster than HNSW, and it
is 2.2--10.5$\times$ faster under adverse correlation.

%% file: 1-intro/intro.tex
Vector queries arrive with predicates on metadata: a product search
wants items in stock and under a price, an image search wants images
with a license, and retrieval-augmented
generation~\cite{rag} wants only the documents its user may read. Filtered vector search answers such a query
with the approximate nearest neighbors among the vectors that pass the
predicate~\cite{fvs-survey}. Four search systems already run it with
\acornone{}~\cite{acorn} on the HNSW~\cite{hnsw} graphs they
serve~\cite{lucene,weaviate,qdrant,vespablog}. The
fraction that passes, the selectivity $\sel$ of the predicate, is often small: 70\% of
the tag queries of the Big ANN filtered track select less than 1\% of
YFCC~10M~\cite{bigannfiltered}, and on LAION~\cite{laion400m} fewer than 1\% of
the images carry a Creative Commons license. This paper targets
extreme selectivity, from 3\% down to 0.1\% (\S\ref{sec:cost}).

Two kinds of predicate are hard for a predicate-aware index to serve. Range and
label
indexes build the predicate into the graph or its posting lists, and they
beat a search over a graph that knows nothing of
it~\cite{serf,irangegraph,filtereddiskann,ung}. Such an index, however,
cannot serve an arbitrary predicate and must be updated when an indexed
attribute changes. A per-query allowlist has no attribute to index, and stock, price
and access rights change after the index is built. A predicate-agnostic method
evaluates the predicate at query time on one graph built without metadata,
so it has neither limitation (\S\ref{sec:landscape}).

Below a few percent selectivity the predicate-agnostic family breaks down,
and its members fail in two opposite ways (\S\ref{sec:problem}).
\acornone{} takes its candidates only from the passing nodes in the one-hop
and two-hop neighborhoods of the node it expands. Its recall therefore
declines as selectivity falls, and on the graphs of this paper it collapses
below 1\% (\S\ref{sec:collapse}, Table~\ref{tab:collapse}). NaviX~\cite{navix}, which adapts
its expansion to the local selectivity, also takes only passing nodes and does
not reach 0.9 from 1\% down (\S\ref{sec:eval-sweep}). HNSW measures every node it
visits and never runs out of candidates, but it visits about $\ef/\sel$
nodes before $\ef$ of them pass. When the passing set is given,
pre-filtering is faster still below about 0.5\%, but the set has to come
from evaluating the predicate on every key or
from an index over attributes chosen in advance
(\S\ref{sec:landscape}).

Both failures follow from what each method adds to its candidate queue,
and that is what we change. \racorn{} extends ACORN's search with a
fallback that keeps it connected at low selectivity, and it leaves
construction untouched. \racornone{} is \racorn{} on the pruned HNSW a system already holds,
the graph \acornone{} runs on in deployed systems (\S\ref{sec:acorn}). The fallback, Adaptive
Search Fallback (ASF), acts on each expansion. When the two-hop expansion
returns fewer passing nodes than a target set by a bridge budget $\br$, ASF
adds failing nodes to the candidate queue as \emph{bridges}, but never adds
them to the result queue. A bridge is expanded like any other
candidate, so a chain of bridges crosses a failing region to passing nodes
more than two hops away, and stride sampling spreads the bridges so that
they do not follow one subtree. To compare \racornone{} with HNSW at equal
time, deadline-based early termination (DET) stops a search whose result
queue is full once the deadline passes. \racornone{} fills its queue long
before HNSW does and spends the remaining time converging, so under a
tight deadline it returns the higher recall.

Our contributions are as follows.
\begin{itemize}
\item Adaptive Search Fallback (\S\ref{sec:asf}--\ref{sec:algorithm}): the
bridge rule, the per-expansion budget that bounds it, stride sampling, and
Algorithm~\ref{alg:racorn} with the added lines marked. Unlike
NaviX, ASF adds failing
nodes to the candidate queue, so a chain of bridges can reach past two hops
(\S\ref{sec:landscape}).
\item Recovery of the recall \acornone{} loses. From 1\% selectivity down
\acornone{} falls short of 0.9 at every $\ef$, and at 0.1\% its result queue
never fills. \racornone{} holds 0.95 or more from 3\% down
(\S\ref{sec:eval-sweep}), and reaches 0.9 on the natural filters and the
predicates beyond a single attribute that \acornone{} misses
(\S\ref{sec:eval-natural}, \S\ref{sec:eval-inexpressible}).
\item An evaluation against the predicate-agnostic family and pre-filtering
on eight indexes of 1M to 100M vectors (\S\ref{sec:eval}). At recall 0.9
\racornone{} is 1.3--8.4$\times$ faster than HNSW at 1\%
selectivity, and it is 2.2--10.5$\times$ faster under adverse correlation.
\item Higher recall at equal time (\S\ref{sec:deadline},
\S\ref{sec:eval-det}). At the same deadline, the time HNSW at $\ef = 100$
or $200$ needs to fill its result queue, \racornone{} has the higher recall
on all datasets and selectivities, at measured latencies equal to HNSW's.
\end{itemize}
The code that reproduces the experiments is at \url{https://github.com/naver/racorn}.

%% file: 2-background/background.tex
Graph-based ANN methods answer a query by traversing a proximity graph and
stepping at each node to the neighbor that lies nearest the query. Following
\acornone{}~\cite{acorn}, we instantiate and evaluate our work on
HNSW~\cite{hnsw}. The mechanism of \S\ref{sec:racorn} reads only the
base-layer adjacency, so it also carries over to flat graphs such as Vamana, the
graph introduced in DiskANN~\cite{diskann} and derived from NSG~\cite{nsg}.

\subsection{HNSW}\label{sec:hnsw}

HNSW organizes the data as a hierarchy of proximity graphs. A node is inserted
into layers $0$ to $\ell$ for a randomly drawn $\ell$. Construction searches
each layer for its $\ef_c$ nearest nodes and links it to $M$
of them, or to $\Mzero = 2M$ at the base layer by convention.
A diversity heuristic chooses them and prunes neighbors dominated by a closer
one. Two degree quantities follow and are not interchangeable: $M$ is the
bound construction enforces, while $\Mavg$ is the average out-degree the graph
actually ends up with at the base layer. The diversity heuristic leaves $\Mavg$
well below the base-layer bound $\Mzero$.

Search descends the upper layers greedily, one candidate at a time, and
switches to a best-first traversal at the base layer. That traversal keeps a
visited set $V$, a candidate queue $C$ ordered by distance to the query, and a
result queue $W$ capped at $\ef$. It repeatedly extracts the nearest candidate
$c$ from $C$, and for each unvisited neighbor $e$ it marks $e$ visited,
computes its distance, and adds it to $C$ when $e$ is closer than the
farthest element $f$ of $W$. The loop ends at the \emph{stopping rule}: $W$
holds $\ef$ entries and $c$ is farther from the query than $f$, so none of
the candidates still in $C$ can enter $W$.

Filtered search over a predicate $F$ takes the form we call \emph{HNSW
in-filtering}, hereafter simply HNSW, which is the behavior of
FAISS~\cite{faiss}, hnswlib~\cite{hnswlib} and USearch~\cite{usearch}. The traversal above is
unchanged except at one point: $e$ enters $W$ only when $F(e)$ holds, while it
still enters $C$ and still costs a distance computation. Two properties follow
and both matter later. First, a distance is computed at every visited node
regardless of the predicate, so the predicate never saves work. Second, $W$
fills with \emph{passing} nodes, so the stopping rule waits for $\ef$ of them
rather than $\ef$ nodes.
When the predicate is independent of the query, a fraction $\sel$ of the
expanded nodes pass, so the search ends after about $\ef/\sel$ expansions as
long as the graph holds at least $\ef$ passing nodes. \S\ref{sec:cost}
measures the distribution, and \S\ref{sec:eval-corr} measures the effect of
correlation.

\subsection{ACORN-1}\label{sec:acorn}

ACORN~\cite{acorn} treats filtered search as traversal over the subgraph
induced by the passing nodes, without building an index for the predicate. \acornone{}, the variant this work extends, leaves construction
untouched: it takes an HNSW graph that exists already and changes only
which nodes the search adds and expands. Build time and memory are therefore
those of the underlying HNSW, so the approach can be deployed on an index a
system is already serving.

The original paper and its adopters, however, run it on different HNSW
graphs, and the choice changes the baseline. ACORN's \S5.3
specifies construction as the ``original HNSW index without pruning'', that
is, with the diversity heuristic disabled. The products that adopted the
method did not follow that specification. Lucene~\cite{lucene},
Weaviate~\cite{weaviate}, Qdrant~\cite{qdrant} and Vespa~\cite{vespablog}
all run it over the pruned graphs they already had rather than grow them.
Omitting the pruning also costs recall: NaviX reports that an unpruned
\acornone{} index needed twice the degree to reach its target recall and
trailed the pruned graph even then~\cite{navix}. We follow the products
throughout. Here \acornone{} means filter-first traversal over a pruned HNSW,
which is also the graph \racornone{} extends, so both methods are measured on
the index a system would actually hold.

\subsection{Filter-First Traversal}\label{sec:filterfirst}

\acornone{} reverses the order of predicate evaluation and distance
computation. At a node $u$ taken from $C$, the predicate is evaluated on the
neighbors of $u$ before any distance is computed, and only passing neighbors
are measured and added to $C$ and $W$. Failing neighbors cost one predicate
evaluation each and nothing more. In particular, they are not marked visited,
so a later expansion that reaches them through a different node may evaluate
them again. Evaluating the predicate first makes the method cheap at 3\%
selectivity and above, but, as \S\ref{sec:collapse} shows, its recall
collapses below 1\%.

One hop alone supplies too few passing candidates once $\sel$ is small, so
ACORN extends the expansion to the two-hop neighborhood. At a visited node it expands
the whole two-hop neighborhood, filters it and truncates the result to $M$,
and it does so at every layer, the upper layers included~\cite{acorn}.

Our \acornone{} runs on the pruned graph of \S\ref{sec:acorn} and keeps the
filter-first rule and the two-hop expansion, with two differences. The upper
layers are descended without the predicate and only the base layer filters,
as Lucene, Weaviate and Vespa do~\cite{lucene,weaviate,vespablog}. The
expansion aims at a target of $\Mzero$ passing candidates and fills it in a
fixed order: it takes the passing one-hop neighbors first; then, when they
number at most one, it gathers the two-hop candidates of all neighbors,
filters them and keeps as many as the target allows, and otherwise it expands
the one-hop neighbors of $u$ in turn, failing ones first, passing ones next
and visited ones last, and stops once the
target is met. At the selectivities this paper studies the target is
rarely met, so our expansion examines the same neighborhood as ACORN's. We write
$|\mathrm{expand}(u)|$ for what the expansion returns, the passing one-hop and
two-hop candidates together; it is at most $\Mzero$ by construction, and
\S\ref{sec:cause} uses it.

The products that adopted \acornone{} each implement the expansion in their
own way~\cite{lucene,weaviate,qdrant,vespablog}. ASF (\S\ref{sec:racorn}) reads
only what an expansion returns and the failing nodes it evaluated, so it
plugs into any of them.

%% file: 3-problem/problem.tex
\subsection{The ACORN-1 Collapse Point}\label{sec:collapse}

\acornone{}'s recall falls as selectivity falls, and below one percent the
decline becomes a collapse. Table~\ref{tab:collapse} reports \acornone{} and
HNSW on three datasets that span 1M to 100M vectors and 128 to 960
dimensions. Each dataset has its own pruned HNSW index, searched at
$\ef=100$ and $k=100$ under a predicate that selects a uniform random $\sel$
fraction of the keys (\S\ref{sec:eval}). Both methods answer the same queries
on the same graph, and they differ only in what they add to the candidate queue.

Down to 3\% selectivity the two sit at opposite ends of a conventional
speed--accuracy trade, and \acornone{} gives up recall for latency. Below 1\%
the trade no longer holds. \acornone{}'s recall falls to near zero at
0.1\%, which is no answer at all. HNSW keeps its recall, but its cost grows
instead. Its cycle count follows $\ef/\sel$ (\S\ref{sec:hnsw}), and its
latency rises with it into hundreds of milliseconds.

The $<k$ column says what the recall column cannot. \acornone{}'s search
does not end at the stopping rule of \S\ref{sec:hnsw}. It ends because the
expansion returns nothing and the candidate queue empties. When the queue
empties before $W$ is full, the query returns fewer than $k$ results. The share
of such queries is 2.4--10.5\% at 1\% selectivity, \mbox{12.8--48\%} at 0.5\% and
all of them at 0.1\%, and the latency falls with it.
\S\ref{sec:cost} counts the cycles behind these shares.

Widening the result queue does not recover the recall. When $\ef$ is swept from 100
to 1600 at 0.5\%, \acornone{} reaches at most 0.53 on each of the three
datasets (\S\ref{sec:eval-sweep}). The binding constraint is the supply of
candidates, not the size of the result queue.

\begin{table}[t]
\caption{\acornone{} and HNSW under a uniform random predicate.
$<k$ is the share of queries (\%) that return fewer
than $k$ results, zero for HNSW throughout, and \emph{ms} is median query
latency.}
\label{tab:collapse}
\centering
\footnotesize
\input{3-problem/figs/collapse_table}
\end{table}

\subsection{Structural Cause}\label{sec:cause}

\begin{table}[t]
\caption{Datasets. One pruned HNSW index per dataset ($M = 16$,
$\ef_c = 100$); $\Mavg$ is its measured base-layer out-degree.}
\label{tab:datasets}
\centering
\scriptsize
\begin{tabular}{@{}l l r l l@{}}
\toprule
Dataset & $N$ & $d$ & metric & $\Mavg$ \\
\midrule
SIFT~\cite{sift,bigann1b} & 1M, 100M & 128 & L2 & 19.30, 21.80 \\
GIST~\cite{sift} & 1M & 960 & L2 & 9.88 \\
YFCC~\cite{yfcc,bigannfiltered} & 1M, 10M & 192 & L2 & 18.43, 18.65 \\
LAION~\cite{laion400m} & 1M, 25M & 512 & cosine & 15.26, 14.69 \\
Text2Image~\cite{bigann} & 40M & 200 & cosine & 19.30 \\
\bottomrule
\end{tabular}
\end{table}

The collapse follows from a single design decision: \acornone{} adds only
predicate-passing nodes to the candidate queue (\S\ref{sec:filterfirst}). Every
candidate it can expand must come from the passing nodes found in the
one-hop and two-hop neighborhoods of a node it has reached. Both
neighborhoods are fixed by the graph, not by the predicate. As $\sel$ falls,
fewer nodes in them pass, and the expansion returns fewer candidates per
cycle until it returns none.

An expansion would return $\Mzero\sel$ passing nodes from one hop and
$\Mzero^{2}\sel$ from two hops only if the graph were regular and its two-hop
neighbors distinct. A pruned HNSW satisfies
neither: its measured base-layer degree $\Mavg$ lies well below the nominal
bound (Table~\ref{tab:datasets}), and its two-hop neighborhoods overlap. What decides the collapse is the expansion size
$|\mathrm{expand}(u)|$, the number of candidates one expansion returns. It
converges to zero for \acornone{} as $\sel$ falls, so the share of queries in
Table~\ref{tab:collapse} that return fewer than $k$ results rises.

HNSW fails in the opposite way. It computes a distance at every visited node
regardless of the predicate, so its candidate queue never starves. It must, however,
keep visiting until $\ef$ \emph{passing} nodes reach $W$, and that takes on
the order of $\ef/\sel$ cycles. HNSW keeps connectivity at the cost of cycles
and \acornone{} keeps cycles low at the cost of connectivity. Neither behavior is
a tuning failure, and neither is repaired by moving $\ef$. Both follow from
what each method adds to its candidate queue.

\subsection{The Filtered ANNS Landscape}\label{sec:landscape}

Filtered vector search systems differ first in when they evaluate the
predicate~\cite{fvs-survey,vbase}. \emph{Pre-filtering} materializes the passing
set and searches it exactly, at a cost proportional to $N\sel$ and the dimension.
\emph{Post-filtering} runs an unfiltered search and applies the predicate
afterwards, so the search width must grow as $\sel$ falls. \emph{In-filtering}
evaluates the predicate during traversal. HNSW, ACORN~\cite{acorn},
NaviX~\cite{navix} and this work belong to it.

Pre-filtering deserves its own cost breakdown, because at the selectivities
this paper targets it is the method to beat. Its cost splits in two. Measuring
every passing vector costs $O(N\sel d)$. When the set is given, only this
half is charged, and pre-filtering wins outright below roughly half a percent
(\S\ref{sec:eval-sweep}). Enumerating the set is the other half. It evaluates
the predicate on every key or intersects inverted lists over the attribute
values. Those lists index scalars rather than vectors. Like the
predicate-aware indexes of the next paragraphs, however, they ask for the
attributes to be chosen before the query arrives. A system that has not chosen its
attributes in advance, the setting of this paper, is left with the scan. On a
multi-attribute conjunction the cost of the scan grows with $N$, while that
of a graph query does not. \S\ref{sec:eval-inexpressible} measures all three forms: the set
given, scanned, and intersected from attribute lists.

A second axis cuts across the timing of the predicate: whether the index is built with
knowledge of what will be filtered. \emph{Predicate-aware} methods put the metadata into the
structure. SeRF~\cite{serf}, iRangeGraph~\cite{irangegraph} and DIGRA~\cite{digra}
materialize subgraphs for numeric ranges. Filtered-DiskANN~\cite{filtereddiskann},
UNG~\cite{ung} and VecFlow~\cite{vecflow} attach labels to the graph or to
posting lists. \emph{Predicate-agnostic} methods build one
graph with no metadata in it and evaluate the predicate at query time.

Predicate-aware indexes are faster than the predicate-agnostic family on the
workloads they were built for. That speed has two limits, and the first is expressiveness. A range index needs an
attribute it can order, and a label index needs a set it can enumerate. Both
need that attribute or label to be known when the index is built. These
structures cannot represent per-query allowlists from access control or
personalization, substring or regular-expression matches over text,
negations of the form $\mathit{label} \notin S$, or values derived at query
time such as $\mathit{price}/\mathit{rating} < \theta$. A range index covers numeric ranges, a label index covers label
equality and conjunctions of labels, and the predicate-agnostic family takes
all of these and everything in the list above.

The second limit is update cost. An attribute baked into the structure is
stale as soon as its value changes, whether it is an ordering in the graph or
a membership in a posting list. Inventory, price, visibility and access
rights change while the index serves. The cost of a change differs between
indexes. SeRF is append-only and iRangeGraph supports neither insertion nor
deletion, so a changed value cannot be applied in place; DIGRA deletes and
reinserts it~\cite{serf,irangegraph,digra}. On the label side VecFlow is static and Filtered-DiskANN accepts only insertions, while
UNG also accepts deletions and so applies a label change in
place~\cite{filtereddiskann,ung,vecflow}. A predicate-agnostic
method reads the current value at query time, so a metadata change costs nothing
on the index side.

A predicate-aware index outperforms the
predicate-agnostic family on the range and label predicates it is built for. We do not compete there, and we
do not compare against such indexes. The advantage is structural, not
incidental. Knowing the predicate at build time lets the index materialize a
subgraph in which every node satisfies it. The graph the query walks is then
navigable by construction, and every distance it computes is spent on a
candidate answer. \racorn{} walks a graph
built without knowledge of the predicate and restores the broken paths by routing through
nodes that fail the predicate, and the distances spent on those detours never
enter the result queue.

The trade is build-time knowledge against query-time traversal cost. When the predicate cannot be known at build time, that
knowledge does not exist for either side, and the predicate-aware index does not
run at all. The per-query allowlist of \S\ref{sec:eval-inexpressible} is
that case.

Our scope follows from this trade. We target workloads where the predicate is
arbitrary and evaluated per query, where the graph already exists and was built
without filter knowledge, and where the metadata changes without reindexing.
None of these three is particular to \racorn{}: they are properties of the
predicate-agnostic family, and HNSW, \acornone{} and NaviX share
them. The family is the only option in these workloads, and
\S\ref{sec:collapse} shows that it has been unusable in exactly the regime
this paper targets. \racorn{} makes it usable there.

NaviX~\cite{navix} is the closest predicate-agnostic work and deserves an exact
comparison. It adapts per candidate: it reads the local selectivity of a
node's neighborhood to choose among expansion heuristics. It uses
predicate-failing neighbors in two ways. They are intermediate nodes that
two-hop expansion passes through, which ACORN already does, and in one
heuristic they are distance computations that order that expansion. None of
the heuristics NaviX uses lets a predicate-failing node enter the candidate queue.

That is where this work differs. Adaptive Search
Fallback (ASF, \S\ref{sec:asf}) adds failing nodes to the candidate queue under an
explicit per-expansion bridge budget, but it keeps them out of the result
queue. A bridge is therefore itself expanded, its expansion can produce
another bridge, and the traversal reaches passing regions more than two hops
away. The two mechanisms are orthogonal. NaviX chooses which passing
neighbors to expand, and ASF decides what to do when there are none. ASF can
therefore be layered on NaviX's choice, but we do not evaluate that
combination.
\S\ref{sec:eval-sweep} compares against a reimplementation of NaviX's adaptive-local
heuristic on the same graph.

\subsection{Design Requirements}\label{sec:requirements}

Two requirements follow.

\begin{itemize}
\item \textbf{R1 (connectivity).} When the passing one-hop and two-hop supply is
exhausted, the traversal needs a detour: a path through nodes that fail the
predicate, which never lets them into the result queue.
\item \textbf{R2 (diversity).} The detours must not concentrate on one
neighborhood: a detour is worth its cost only if it reaches a region the
traversal does not already see. Taking them from the head of a sorted list
would follow a single one-hop neighbor's subtree.
\end{itemize}

\S\ref{sec:asf} meets R1 with Adaptive Search Fallback and \S\ref{sec:stride}
meets R2 with stride sampling.

%% file: 3-problem/figs/collapse_table.tex
\begin{tabular}{@{}l r r r r r@{}}
\toprule
 & \multicolumn{3}{c}{ACORN-1} & \multicolumn{2}{c}{HNSW} \\
\cmidrule(lr){2-4}\cmidrule(lr){5-6}
$s$ & recall & $<k$ & ms & recall & ms \\
\midrule
\multicolumn{6}{@{}l}{\textit{GIST 1M (960d)}} \\
3\% & 0.628 & 1.1 & 2.7 & 0.908 & 65.3 \\
1\% & 0.379 & 10.5 & 1.5 & 0.936 & 138.8 \\
0.5\% & 0.147 & 48.0 & 1.0 & 0.948 & 226.7 \\
0.1\% & 0.005 & 100 & 0.1 & 0.965 & 590.2 \\
\midrule
\multicolumn{6}{@{}l}{\textit{SIFT 100M (128d)}} \\
3\% & 0.669 & 0.3 & 6.0 & 0.951 & 38.9 \\
1\% & 0.403 & 2.4 & 5.8 & 0.970 & 100.9 \\
0.5\% & 0.171 & 12.8 & 6.0 & 0.977 & 185.1 \\
0.1\% & 0.005 & 100 & 0.1 & 0.989 & 754.6 \\
\midrule
\multicolumn{6}{@{}l}{\textit{LAION 25M (512d)}} \\
3\% & 0.729 & 0.9 & 3.6 & 0.955 & 30.0 \\
1\% & 0.409 & 9.6 & 3.7 & 0.960 & 77.2 \\
0.5\% & 0.120 & 43.2 & 3.2 & 0.963 & 145.5 \\
0.1\% & 0.003 & 100 & 0.1 & 0.969 & 616.7 \\
\bottomrule
\end{tabular}

%% file: 4-racorn/racorn.tex
\racornone{} is \acornone{} with one addition to the expansion step.
Construction is untouched, the filter-first rule and the two-hop
expansion of \S\ref{sec:filterfirst} run as before, and only passing nodes still
enter the result queue. The addition acts when an expansion returns too
little. Adaptive Search Fallback (\S\ref{sec:asf}) adds a bounded number of
failing nodes to the candidate queue as \emph{bridges}, and so meets R1.
Stride sampling (\S\ref{sec:stride}) chooses which failing nodes become
bridges, and so meets R2. Algorithm~\ref{alg:racorn} (\S\ref{sec:algorithm})
gives the procedure with the added lines marked, and \S\ref{sec:cost} gives its
cost.

\input{4-racorn/figs/asf-bridge}

\subsection{Adaptive Search Fallback}\label{sec:asf}

ASF works per expansion and asks one question of it: did the expansion supply
enough? Let $n$ be the number of unvisited one-hop neighbors of the node $u$
being expanded, $C_1$ and $C_2$ the passing nodes it returns from one and two hops,
$F_1$ and $F_2$ the failing ones, and $P = C_1 \cup C_2$ the candidates the
expansion returns, at most $\Mzero$ of them (\S\ref{sec:filterfirst}). ASF sets
the expansion a target $\tau = \lfloor \br\, n \rfloor$, where $\br$ is the
bridge budget, the one parameter \racornone{} adds. ASF acts when the
expansion falls short of that target, $|P| < \tau$. It then fills the expansion up to $\tau$ with nodes
drawn from $F_2$. Each of them is measured against the query and pushed onto
the candidate queue $C$, but none enters $W$. These are the bridges. The fill
happens only while $W$ holds fewer than $\ef$ entries. Once the result queue
is full, expansions return passing nodes only, as in \acornone{}.

A bridge is a candidate, not a result. It is expanded like any other
candidate when it is popped from $C$: its passing neighbors enter $W$, and if
its own expansion falls short, that shortfall is filled with bridges again.
This is the detour R1 asks for, and Fig.~\ref{fig:asf} contrasts it with
what \acornone{} does on the same neighborhood. There the one-hop and two-hop neighborhoods of $u$ hold no passing node, so
\acornone{}'s expansion returns nothing and its candidate queue empties, while
a path of bridges crosses the failing region and arrives at a passing node. Such a
path can chain any number of bridges while $W$ is filling. Because a bridge never
enters $W$, every node returned passes the predicate, and
the distance computed for a bridge serves only to order $C$ and so decides
when the bridge is expanded.

On a cycle where ASF fires, every failing node the expansion touched is marked
visited, whether or not it was chosen as a bridge. \acornone{} leaves failing
nodes unmarked, so a failing node is evaluated again from every node that
reaches it. The repeated evaluations are rare while few nodes fail, but at low
selectivity most of a neighborhood fails and the same nodes are evaluated over
and over. Marking them ends the repetition. It also means that a
failing node not chosen as a bridge on a firing cycle is not offered again.
\S\ref{sec:stride} describes how the bridges are chosen.

We scale the target with the unvisited degree $n$ of the node rather than
fixing it at $\Mzero$, for two reasons. First, a node whose neighbors are mostly visited already sits in
explored territory and should draw few bridges, and the target shrinks with
$n$ accordingly. Second, $\br$ then keeps one meaning across graphs of
different density. At $\br = 1$ the target is $n$, so a short expansion, one with
$|P| < \tau$, is filled up to $n$ nodes, passing candidates and bridges together. That is what
the node would have supplied on its own if all $n$ unvisited neighbors had
passed the predicate. At $\br = 2$ the target is $2n$. As $\br \to 0$ no bridge is ever added and the procedure is \acornone{}. A
larger $\br$ adds more failing nodes per short expansion, at one distance
computation each. The target counts every node the expansion adds, bridges
included, so $\Mzero$ caps $P$ but not $\tau$. At $\br > 1$ a short expansion
can therefore add more than $\Mzero$ nodes. \S\ref{sec:cost} measures what
$\br$ costs. \S\ref{sec:eval} runs at one fixed $\br$, and
\S\ref{sec:eval-det} and \S\ref{sec:eval-ablation} vary it.

As a result, the candidate queue does not run dry while $W$ is filling. \acornone{}'s queue drains when expansions return
nothing, which at 0.1\% happens almost at once (\S\ref{sec:cost}). Under ASF
an expansion that returns no passing node still adds $\tau$ bridges whenever
the failing two-hop pool $F_2$ holds at least $\tau$ nodes, and at low
selectivity that pool is nearly the whole two-hop neighborhood. The traversal therefore ends at
the stopping rule with $W$ full rather than at exhaustion with $W$ nearly
empty. \S\ref{sec:cost} reports the cycles this adds, and \S\ref{sec:cost}
and \S\ref{sec:eval} measure how much of the added traversal turns into
recall.

Two properties distinguish ASF. First, ASF needs only two inputs, the
base-layer neighbor lists of the expanded node and the predicate. The upper layers only choose the
entry point, so the mechanism is not tied to the hierarchy. Second, ASF differs
from deeper expansion. Vespa offers this as an option: a third hop when the
two-hop expansion runs short, with distances still computed only for passing
nodes~\cite{vespablog}. \S\ref{sec:eval-sweep}
measures a three-hop \acornone{} against ASF. A third hop stops only once it
has found $\Mzero$ passing candidates, which a three-hop neighborhood rarely
holds at low selectivity. So each short expansion while $W$ is filling
examines most of that neighborhood, and because failing nodes stay unmarked,
the next cycle examines much of it again. It also adds whatever passes within its radius, regardless of
direction. A bridge instead costs one distance computation and is expanded
only when it is the nearest candidate left in $C$. The traversal therefore
moves toward the query through failing regions.

\subsection{Stride Sampling}\label{sec:stride}

The choice of failing nodes as bridges matters as much as their number, and
the pool they are drawn from is not in a neutral order. The two-hop expansion takes the
one-hop neighbors of $u$ in turn and appends the failing two-hop nodes it finds
under each, so $F_2$ is a sequence of blocks, one per one-hop neighbor. Taking
the first $\tau - |P|$ entries would draw every bridge from the subtree of the
first neighbor or two, and the traversal would push in one direction. This is
the concentration R2 is meant to prevent.

Stride sampling spreads the bridges across the blocks at no extra cost. To choose
$m$ bridges from a pool of $|F_2|$ nodes it keeps every $\lfloor |F_2|/m
\rfloor$-th entry, so the selection reaches nearly every block whose length is at least the
stride and takes at most a few bridges from each. The selection is by position in
the list and computes no distance. Each chosen bridge then costs one distance
computation, as every bridge does.

When more two-hop nodes pass than the target $\Mzero$ allows
(\S\ref{sec:filterfirst}), the same sampler is applied, so the expansion
returns a spread sample rather than a prefix. At the selectivities this paper
targets, that case is rare, so the effect that \S\ref{sec:eval-ablation}
reports against prefix truncation is the effect on bridges.

\subsection{Algorithm}\label{sec:algorithm}

\begin{algorithm}[t]
\caption{\racornone{} base-layer traversal. Lines 1--11 are \acornone{} as
\S\ref{sec:filterfirst} defines it; lines 12--18 are ASF.}
\label{alg:racorn}
\begin{algorithmic}[1]
\REQUIRE query $q$, predicate $F$, entry node $p$, result size $\ef$, bridge budget $\br$
\ENSURE $W$, the $\ef$ nearest passing nodes found
\STATE $C \gets \{p\}$;\; $V \gets \{p\}$;\; $W \gets \{p\}$ if $F(p)$, else $\emptyset$
\WHILE{$C \neq \emptyset$}
  \STATE $u \gets$ pop the nearest node from $C$
  \IF[$f$: farthest in $W$]{$|W| = \ef$ and $d(q,u) > d(q,f)$}
    \STATE \textbf{break}
  \ENDIF
  \STATE $N_u \gets$ unvisited neighbors of $u$;\; $n \gets |N_u|$
  \STATE $C_1 \gets \{v \in N_u : F(v)\}$;\; $F_1 \gets N_u \setminus C_1$ \COMMENT{predicate only}
  \STATE $(C_2, F_2) \gets \textsc{ExpandTwoHop}(u, \Mzero - |C_1|)$
  \STATE $P \gets C_1 \cup C_2$ \COMMENT{$|P| \le \Mzero$}
  \FOR{$v \in P$}
    \STATE $V \gets V \cup \{v\}$; compute $d(q,v)$; push $v$ onto $C$, and onto $W$ if $|W| < \ef$ or $d(q,v) < d(q,f)$, evicting $f$ when $|W|$ exceeds $\ef$
  \ENDFOR
  \STATE $\tau \gets \lfloor \br\, n \rfloor$ \COMMENT{ASF target}
  \IF{$|P| < \tau$}
    \STATE $V \gets V \cup F_1 \cup F_2$ \COMMENT{no re-evaluation later}
    \IF{$|W| < \ef$}
      \STATE $B \gets \textsc{StrideSample}(F_2, \tau - |P|)$ \COMMENT{\S\ref{sec:stride}}
      \FOR{$v \in B$}
        \STATE compute $d(q,v)$; push $v$ onto $C$ \COMMENT{not $W$}
      \ENDFOR
    \ENDIF
  \ENDIF
\ENDWHILE
\RETURN $W$
\end{algorithmic}
\end{algorithm}

Algorithm~\ref{alg:racorn} gives the base-layer traversal of \racornone{},
starting from the entry node $p$ selected by the greedy descent through the
upper layers. Lines 1--11 are the procedure of \S\ref{sec:filterfirst},
and lines 12--18 are what ASF adds. We walk through one cycle.

A cycle takes the nearest candidate $u$ from $C$ (line 3) and applies the
stopping rule of \S\ref{sec:hnsw} (line 4): once $W$ holds $\ef$ nodes and $u$
is farther than the farthest of them, every remaining candidate is farther than
$u$, so $W$ is final.
The one-hop scan (lines 6--7) evaluates the predicate on the unvisited
neighbors of $u$ and splits them into $C_1$ and $F_1$. No distance is
computed here, and this is the filter-first rule of \acornone{}. The two-hop
expansion (line 8) reads the neighbor lists of the neighbors of $u$ and returns the passing
two-hop candidates $C_2$ and the failing pool $F_2$. It aims at $\Mzero - |C_1|$ passing ones, so that the
expansion as a whole returns at most $\Mzero$. The passing candidates are then
measured and added (lines 10--11). Each enters $C$, and it also enters $W$
when $W$ has room or $v$ is closer than the current farthest result. $W$
holds at most $\ef$ entries.

ASF begins at line 12 with the target $\tau$. If the expansion fell short of
it (line 13), the failing nodes of both hops are marked visited (line 14). While $W$ is not yet full (line 15),
the shortfall $\tau - |P|$ is then drawn from $F_2$ by stride sampling
(line 16). Each bridge is measured and pushed onto $C$ alone (line 18). When
$C$ later pops it, it is expanded by the same lines 6--18 as any other
node, so a bridge can lead to another bridge.

The added lines change neither the stopping rule nor the construction, and
they act only on cycles whose expansion fell short of the target. At
selectivities where expansions are rarely short, they rarely run, and
\racornone{} performs the cycles \acornone{} does (\S\ref{sec:cost}).

%% file: 4-racorn/figs/asf-bridge.tex
\begin{figure}[t]
\centering
\begin{tikzpicture}[
  x=7.6mm, y=6.4mm,
  pass/.style={circle, draw, fill=black, inner sep=0pt, minimum size=2.0mm},
  fail/.style={circle, draw, fill=white, inner sep=0pt, minimum size=2.0mm},
  cur/.style={circle, draw, fill=black, inner sep=0pt, minimum size=2.9mm},
  edge/.style={gray!60, line width=0.35pt},
  admit/.style={-{>}, line width=0.8pt},
  bridge/.style={-{>}, line width=0.8pt, densely dashed},
  lbl/.style={font=\footnotesize, inner sep=1pt},
]
  \fill[black!8, rounded corners=2pt] (7.0,-1.6) rectangle (9.8,1.6);
  \node[lbl, text=black!55] at (8.4,2.1) {$F$ holds};
  \foreach \p/\x/\y in {a/1.7/1.1, b/1.7/-1.1, c/3.3/1.6, d/3.3/0.0,
                        e/3.3/-1.6, f/5.0/1.1, g/5.0/-1.1, h/6.2/0.0}
    \node[fail] (\p) at (\x,\y) {};
  \foreach \p/\x/\y in {p1/7.6/0.8, p2/9.0/1.1, p3/8.2/-0.4, p4/9.3/-1.1}
    \node[pass] (\p) at (\x,\y) {};
  \node[cur, label={[lbl]below:$u$}] (u) at (0.2,0) {};
  \foreach \s/\t in {u/a, u/b, a/c, a/d, b/d, b/e, c/f, d/f, d/g, e/g,
                     f/h, g/h, h/p1, h/p3, p1/p2, p3/p4, p1/p3, p2/p4}
    \draw[edge] (\s) -- (\t);
  \draw[gray, densely dotted, rounded corners=3pt] (1.0,-2.2) rectangle (4.1,2.2);
  \node[lbl, text=black!55] at (2.55,2.6) {\acornone{} examines};
  \draw[bridge] (u) -- (d);
  \draw[bridge] (d) -- (h);
  \draw[admit] (h) -- (p1);
  \draw[admit] (h) -- (p3);
  \begin{scope}[shift={(0.0,-3.3)}]
    \node[pass] (lp) at (0,0) {};
    \node[lbl, right=1.0mm of lp] (lpt) {passes $F$};
    \node[fail, right=5mm of lpt] (lf) {};
    \node[lbl, right=1.0mm of lf] (lft) {fails $F$};
  \end{scope}
  \begin{scope}[shift={(0.0,-4.2)}]
    \draw[admit] (0,0) -- ++(7mm,0)
      node[lbl, right=0.8mm] (lat) {added to $C$ and $W$};
    \draw[bridge] ([xshift=6mm]lat.east) -- ++(7mm,0)
      node[lbl, right=0.8mm] {bridge: $C$ only};
  \end{scope}
\end{tikzpicture}
\caption{Adaptive Search Fallback at low selectivity, on one base-layer
neighborhood: the node $u$ being expanded, a region where every node fails
the predicate, and a passing region several hops away. The dotted
outline is all \acornone{} examines, the one-hop and two-hop neighborhood of
$u$; it holds no passing node, so the expansion returns nothing and the
candidate queue empties. The dashed arrows are what ASF adds: failing nodes
that enter the candidate queue as bridges, under the bridge budget. Each bridge
is two hops from the node whose expansion added it, so the chain crosses the
failing region and reaches the passing one. Bridges never enter the result
queue $W$.}
\label{fig:asf}
\end{figure}

%% file: 4-racorn/cost.tex
This section compares \acornone{}, HNSW and \racornone{} at the same $\ef$
on the eight indexes of Table~\ref{tab:datasets}, with \racornone{} at
$\br = 1$ and $2$ (Tables~\ref{tab:recall} and \ref{tab:cost}). Unless stated
otherwise, the numbers below are ranges over the eight datasets at $\br = 2$,
the setting of \S\ref{sec:eval}, and $\ef = 100$. At equal $\ef$
HNSW has the higher recall on every cell (each dataset at each selectivity),
so the comparison is one of cost at unequal recall. From 3\% down the recall gap
is 0.01--0.28 (Table~\ref{tab:recall}), and \racornone{} returns in 5--23$\times$ less time. At
$\ef = 200$ the gap narrows and the time ratios stay
(Table~\ref{tab:cost}). \S\ref{sec:eval-sweep} asks which method reaches a
given recall sooner, and \S\ref{sec:deadline} and \S\ref{sec:eval-det} ask
which returns the better answer in a given time.

\begin{table}[t]
\caption{\racornone{} at $\br = 0$ (\acornone{}), $1$ and $2$, and HNSW, at $\ef = 100$:
recall and base-layer cycles, as ranges over the eight datasets of \S\ref{sec:eval}.}
\label{tab:recall}
\centering
\scriptsize
\input{4-racorn/figs/recall_table}
\end{table}

\begin{table}[t]
\caption{HNSW against \racornone{}, as ranges over the eight datasets: dist.\ and
pred.\ are distance computations and predicate evaluations, latency is the
median over the queries, and $T/$bound is \racornone{}'s cycles over the
right-hand side of~(\ref{eq:bound}).}
\label{tab:cost}
\centering
\scriptsize
\input{4-racorn/figs/cost_table}
\end{table}

At $\br = 0$ \racornone{} is \acornone{} (\S\ref{sec:asf}), and the two
diverge where the two-hop expansion runs short. \acornone{} runs about $\ef$
cycles regardless of selectivity down to 0.5\%. At 0.1\% it runs a single
cycle and no query fills $W$ (Table~\ref{tab:collapse}), and its recall
falls with the supply of passing candidates (Table~\ref{tab:recall}). \racornone{} spends more
cycles as $\sel$ falls, and every query fills $W$, so its recall holds at both $\br = 1$ and $2$ (Table~\ref{tab:recall}).

Above 3\% the difference fades. At 5\% and 10\% ASF fires on 2--5\% of the
cycles, so \racornone{} runs close to \acornone{} (the 5\% row of
Table~\ref{tab:recall}). From 3\%
down the bridges carry the search, the cycle count grows with $1/\sel$ (Table~\ref{tab:recall}), and
the bound~(\ref{eq:bound}) below holds on every cell at both $\br$. This is why both the
bound and the range this paper targets start at 3\%.

A cycle is not the same work in each method. We count the cycles of the
base-layer traversal, one per candidate popped from $C$, and leave out the
descent through the upper layers. By itself, a cycle adds only the queue and
neighbor-list work of expanding one node, whose distance was computed when
it was pushed. HNSW computes a distance for every unvisited neighbor of the
node it expands, up to $\Mavg$ per cycle, and evaluates the predicate on
those that pass the radius test and enter $C$. \racornone{} first evaluates the predicate on the one-hop and
two-hop neighborhood and computes distances only for the passing nodes and
the bridges. In both methods the stopping rule leaves the rest of the
queue unexpanded: those candidates were measured when pushed, but no cycle is
spent on them. Cycles, distance computations and predicate evaluations
are therefore three different counts, and the time is roughly the sum of the
three, each weighted by its unit cost.

HNSW spends about $\ef/\sel$ cycles (\S\ref{sec:hnsw}). The ratio of cycles
to $\ef/\sel$ is the center of a per-query distribution, not an identity.
Its median lies in 0.96--1.02 on every cell, and the 1st to 99th
percentiles lie within 0.76--1.54. The relation also needs $N\sel \ge \ef$:
at $N\sel = \ef/2$ the queue empties before $W$ fills, and the ratio falls
to about half. \racornone{}'s cycle count satisfies
\begin{equation}
  T_{\racornone} \le \frac{\ef/\sel}{\Mavg}\,\br
  \;\;\text{for } \br \ge 1,\ \sel \le 3\%\ (M = 16),
  \label{eq:bound}
\end{equation}
with the predicate independent of the query as in \S\ref{sec:hnsw}, so its
saving over HNSW is at least $\Mavg/\br$.
A rough argument suggests the form of~(\ref{eq:bound}): about $\ef/\sel$
nodes must be touched before $\ef$ of them pass, an expansion touches a two-hop
neighborhood of about $\Mavg^{2}$ nodes, so about $(\ef/\sel)/\Mavg^{2}$
expansions would do if neighborhoods did not overlap, and each can add up to
$\tau \approx \br\Mavg$ bridges that cost a cycle each; their product is the
right-hand side. This is not a proof; the bound is validated empirically below.

The empirical bound held on every cell at $\ef \in \{100, 200\}$ and
$\br \in \{1, 2\}$, with a largest ratio of 0.7 (Table~\ref{tab:cost}). It
also held throughout the $\br$ sweep of \S\ref{sec:eval-ablation}, which
shows how the 3\% moves with $M$ (Fig.~\ref{fig:ablation}c). The bound is
loose because the right-hand side of~(\ref{eq:bound}) lets every cycle add the
full target $\tau$, while ASF adds bridges only while $W$ is filling. At $\br = 1$ the
measured saving in cycles is 25--56$\times$ (Table~\ref{tab:cost}), whereas~(\ref{eq:bound})
guarantees only $\Mavg/\br = \Mavg$, 9.88--21.80$\times$ (Table~\ref{tab:datasets}).

Distance computations fall more than cycles do, by \mbox{20--97}$\times$, because
per cycle \racornone{} measures only passing nodes and bridges. Predicate evaluations rise
instead, by 3--7$\times$ (\mbox{Table~\ref{tab:cost}}), because the two-hop
neighborhood of every expanded node is tested. The saving in time therefore
holds while one predicate evaluation costs less than 0.86--1.57 distance
computations. An attribute comparison costs far below that range, as the
latencies of \S\ref{sec:eval-sweep} confirm, and
\S\ref{sec:eval-inexpressible} includes an expensive predicate.

Raising $\br$ increases recall at the cost of cycles. From $\br = 1$ to $2$
the recall gap to HNSW narrows (Table~\ref{tab:recall}), while cycles and
distance computations rise by 1.2--1.9$\times$. \S\ref{sec:eval-ablation} sweeps $\br$ further.

Both methods spend most of their time converging after $W$ fills. HNSW
fills $W$ in 7--33\% of its latency, and \racornone{} fills it in 9--36\% of
its own. In absolute time, however, \racornone{} fills $W$
4.9--13.2$\times$ sooner and in 17--28$\times$ fewer cycles. While filling,
it evaluates the predicate only slightly more often than HNSW, since both
methods must
test about $\ef/\sel$ nodes before $\ef$ of them pass. The 3--7$\times$ rise
in predicate evaluations builds up after $W$ fills.

Filling $W$ sooner buys time, not a better answer. At the
instant $W$ fills, \racornone{}'s recall is 0.18--0.71 against HNSW's
0.37--0.80, and $\ef = 200$ gives the same picture. Filling early instead
leaves \racornone{} time to converge while HNSW is still filling, and
\S\ref{sec:deadline} turns that time into recall under a deadline.

%% file: 4-racorn/figs/recall_table.tex
\setlength{\tabcolsep}{1.0pt}
\begin{tabular}{@{}l r r r r r r r r@{}}
\toprule
 & \multicolumn{4}{c}{recall} & \multicolumn{4}{c}{cycles} \\
\cmidrule(lr){2-5}\cmidrule(lr){6-9}
$\sel$ & $\br = 0$ & $\br = 1$ & $\br = 2$ & HNSW & $\br = 0$ & $\br = 1$ & $\br = 2$ & HNSW \\
\midrule
5\% & 0.69--0.91 & 0.68--0.91 & 0.68--0.92 & 0.88--0.99 & 101--103 & 103--110 & 112--139 & 2.0k \\
3\% & 0.60--0.85 & 0.61--0.87 & 0.64--0.91 & 0.90--0.99 & 101--103 & 110--134 & 132--194 & 3.3k \\
1\% & 0.33--0.59 & 0.57--0.87 & 0.64--0.94 & 0.93--1.00 & 102--106 & 195--302 & 298--516 & 10k \\
0.5\% & 0.12--0.32 & 0.60--0.91 & 0.68--0.96 & 0.94--1.00 & 101--109 & 357--574 & 598--1.0k & 20k \\
0.1\% & 0.00--0.01 & 0.70--0.97 & 0.78--0.98 & 0.96--1.00 & 1 & 1.8k--3.2k & 3.2k--6.0k & 96k--100k \\
\bottomrule
\end{tabular}

%% file: 4-racorn/figs/cost_table.tex
\setlength{\tabcolsep}{1.5pt}
\begin{tabular}{@{}l r r r r r r r r r r@{}}
\toprule
 & \multicolumn{6}{c}{$\dfrac{\text{HNSW}}{\text{\racornone{}}}$} & \multicolumn{2}{c}{$\dfrac{\text{\racornone{}}}{\text{HNSW}}$} & \multicolumn{2}{c}{\racornone{}} \\
\cmidrule(lr){2-7}\cmidrule(lr){8-9}\cmidrule(lr){10-11}
 & \multicolumn{2}{c}{cycles} & \multicolumn{2}{c}{dist.} & \multicolumn{2}{c}{latency} & \multicolumn{2}{c}{pred.} & \multicolumn{2}{c}{$T/$bound} \\
\cmidrule(lr){2-3}\cmidrule(lr){4-5}\cmidrule(lr){6-7}\cmidrule(lr){8-9}\cmidrule(lr){10-11}
$\sel$ & $\br = 1$ & $\br = 2$ & $\br = 1$ & $\br = 2$ & $\br = 1$ & $\br = 2$ & $\br = 1$ & $\br = 2$ & $\br = 1$ & $\br = 2$ \\
\midrule
\multicolumn{11}{@{}l}{\textit{$\ef = 100$}} \\
5\% & 18--19 & 14--18 & 13--15 & 12--14 & 4--12 & 4--11 & 6--11 & 6--9 & 0.5--1.1 & 0.3--0.6 \\
3\% & 25--30 & 17--25 & 26--31 & 20--25 & 7--22 & 6--18 & 4--6 & 5--7 & 0.3--0.7 & 0.2--0.5 \\
1\% & 33--51 & 20--34 & 60--95 & 38--62 & 10--35 & 7--23 & 3--4 & 3--5 & 0.2--0.5 & 0.2--0.5 \\
0.5\% & 35--56 & 19--34 & 73--137 & 42--83 & 10--33 & 6--19 & 2--3 & 3--5 & 0.2--0.5 & 0.2--0.4 \\
0.1\% & 31--56 & 17--31 & 60--179 & 31--97 & 8--19 & 5--11 & 2--3 & 3--5 & 0.3--0.5 & 0.3--0.5 \\
\midrule
\multicolumn{11}{@{}l}{\textit{$\ef = 200$}} \\
5\% & 18--20 & 14--18 & 13--16 & 13--14 & 4--12 & 4--11 & 6--11 & 5--10 & 0.5--1.1 & 0.3--0.6 \\
3\% & 24--30 & 17--24 & 27--31 & 21--25 & 6--20 & 6--16 & 4--6 & 4--7 & 0.4--0.7 & 0.2--0.5 \\
1\% & 32--49 & 19--31 & 60--93 & 37--60 & 9--29 & 7--19 & 3--4 & 3--5 & 0.3--0.5 & 0.2--0.5 \\
0.5\% & 33--54 & 19--31 & 69--133 & 39--78 & 8--25 & 6--16 & 2--3 & 3--5 & 0.3--0.5 & 0.3--0.5 \\
0.1\% & 26--53 & 14--29 & 44--165 & 23--90 & 6--13 & 4--9 & 2--3 & 2--4 & 0.4--0.6 & 0.4--0.6 \\
\bottomrule
\end{tabular}

%% file: 5-deadline/deadline.tex
In a service where latency matters, the question for a filtered
search is not how fast it finishes but how good an answer it has by a given
deadline. This question sets a third requirement beside R1 and R2 of
\S\ref{sec:requirements}. Within the deadline, the search must spend the
remaining time on candidates that can still change the answer, and it must
stop when the result queue is full rather than when the search converges. Early termination of this
kind has been proposed with a learned predictor as its
trigger~\cite{learned-et,darth}. Here the trigger is the deadline itself, so
that we compare \racornone{} with HNSW at equal time.

\subsection{Deadline-based Early Termination}\label{sec:det}

The stopping rule of \S\ref{sec:hnsw} runs in two phases. The traversal
first fills $W$ with $\ef$ passing nodes and then keeps expanding until the
nearest candidate is farther than the farthest result. In this second phase,
the convergence, $W$ changes only when a closer passing node turns up.
Deadline-based early termination (DET) leaves the first phase alone and cuts
the second. Once $W$ holds $\ef$ entries, the search reads the clock at a fixed
interval of cycles beside the stopping rule of line~4 in Algorithm~\ref{alg:racorn}
and returns at the first check after the deadline. Before $W$ is full the clock is not read, so
the deadline never changes how many neighbors come back, only how close they
are. The deadline is therefore not a bound on latency, since filling
$W$ can take longer than the deadline.

\S\ref{sec:cost} shows what a deadline does to each method. HNSW spends
7--33\% of its time filling $W$ and the rest converging, while from 3\% down
\racornone{} has $W$ full within about a fifth of that time. A deadline
equal to HNSW's own filling time therefore stops HNSW the moment its first
$\ef$ results are in, before any convergence. At that deadline \racornone{} has spent
most of the time converging on a full queue. HNSW cannot hold a shorter
deadline, since it keeps searching until $W$ is full, while a longer one
lets its convergence catch up. \S\ref{sec:eval-det} therefore uses HNSW's
filling time as the reference deadline, fixed by HNSW's own behavior rather
than chosen by us.

\subsection{Scope}\label{sec:det-scope}

Two properties of DET limit the comparison. First, the trigger is wall-clock time, so
the same query on the same index can return a different result under a
different load. The measurements of \S\ref{sec:eval-det} were taken on
dedicated machines. Second, DET changes nothing about which candidates are
expanded, so it trades the tail of the convergence for time.
\S\ref{sec:eval-det} measures what that tail was worth in recall.

%% file: 6-eval/eval.tex
We evaluate on the eight indexes of Table~\ref{tab:datasets}, which span
two orders of magnitude in $N$, from 1M to 100M vectors, dimensions from
128 to 960, hand-crafted descriptors and learned embeddings, and both
distance functions. SIFT~100M is the first 100M vectors of BIGANN~\cite{bigann1b} and the
YFCC vectors are the CLIP~\cite{clip} embeddings distributed with the Big ANN
filtered track~\cite{bigannfiltered}, whose tag queries \S\ref{sec:eval-natural} uses.

The controlled predicate assigns each key a uniform random 32-bit attribute
from a fixed seed and selects the keys below a threshold. The
threshold for selectivity $\sel$ thus selects an $\sel$ fraction of the keys,
at $O(1)$ cost per predicate evaluation and with no correlation to the query.
\S\ref{sec:eval-corr}, \S\ref{sec:eval-natural} and
\S\ref{sec:eval-inexpressible} replace it in turn with a predicate of
controlled correlation, with the attributes and tags the datasets ship with,
and with predicates beyond a single range or label.

Seven methods run on the same index, a pruned HNSW built once per dataset
with $M = 16$ and $\ef_c = 100$, the defaults of Lucene~\cite{lucene-hnsw} and
Qdrant~\cite{qdrant-indexing}, in a fork of USearch. Pre-filtering appears
twice, with the passing set given and with the set obtained by scanning the predicate over
every key (\S\ref{sec:landscape}). \S\ref{sec:eval-natural} and
\S\ref{sec:eval-inexpressible} add the third form, which reads the set from
an index over the attribute. HNSW, \acornone{} and its three-hop variant
(\S\ref{sec:asf}) and \racornone{} are the implementations of
\S\ref{sec:background} and \S\ref{sec:racorn}. NaviX is our reimplementation
of its adaptive-local heuristic on the same base-layer loop. It follows the
graph search of the authors' artifact~\cite{navix-artifact}. The artifact
switches to brute force when at most 30,000 nodes pass and returns nothing at
selectivity 0.1\% or below. We remove both rules and keep the graph search.
The brute force is the same as our pre-filtering with the passing set given. \racornone{} runs at $\br = 2$ with stride sampling on,
except that \S\ref{sec:eval-det} varies $\br$ and \S\ref{sec:eval-ablation} varies both settings. \acornone{}, its
three-hop variant and \racornone{} run behind an entry gate, which evaluates
the predicate on the base-layer entry node and its one-hop and two-hop
neighbors. The query runs filter-first only when at most 30\% of them pass,
and runs as HNSW otherwise. In every cell the median query ran filter-first,
except the 33\% conjunction of \S\ref{sec:eval-inexpressible}, where most
queries went to HNSW.

Every query asks for $k = 100$ neighbors, and each method runs at $\ef$ from
100 to 1600 until it reaches recall 0.97. Each cell is 992
queries, timed over three runs after two warm-ups, and we report the median
latency. QPS is the throughput of 32 search threads bound to one NUMA node
(32 cores) of a two-socket AMD EPYC 7543 server with 1~TB of memory. Every method is compiled with GCC~11.5.0 (\mbox{-O3}
\mbox{-march=native}) and uses the same AVX2 distance kernels from
NumKong~\cite{numkong}. All methods of one dataset run in one process, so
that their latencies are comparable.

\subsection{Selectivity Sweep}\label{sec:eval-sweep}
\input{6-eval/61-sweep}
\subsection{Correlation Robustness}\label{sec:eval-corr}
\input{6-eval/62-correlation}
\subsection{Natural Filters}\label{sec:eval-natural}
\input{6-eval/63-natural}
\subsection{Predicates Beyond a Single Attribute}\label{sec:eval-inexpressible}
\input{6-eval/64-inexpressible}
\subsection{Recall Under a Deadline}\label{sec:eval-det}
\input{6-eval/65-det}
\subsection{Ablations}\label{sec:eval-ablation}
\input{6-eval/66-ablation}

%% file: 6-eval/61-sweep.tex
\newcommand{\figorpending}[2]{\IfFileExists{#1}{\includegraphics[width=\textwidth]{#1}}{\fbox{\parbox{\dimexpr\textwidth-2\fboxsep-2\fboxrule}{\centering\vspace{#2}figure pending\vspace{#2}}}}}

\begin{figure*}[t]
\centering
\figorpending{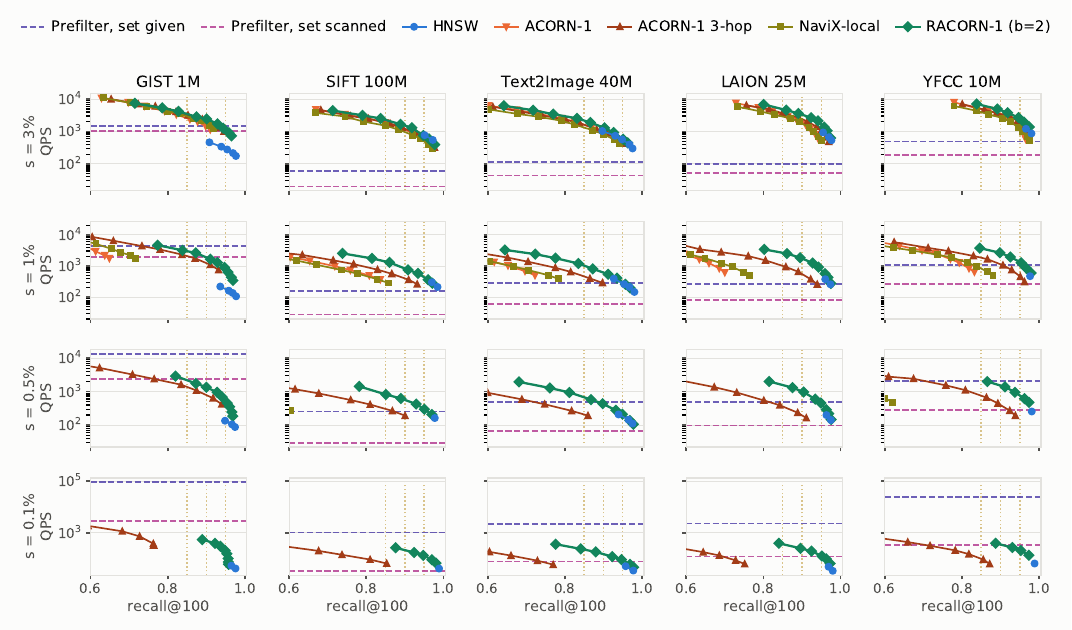}{1.6in}
\caption{QPS against recall on the five datasets of
Table~\ref{tab:datasets}, one
column per dataset and one row per selectivity (3\%, 1\%, 0.5\%, 0.1\%),
$\ef$ swept from 100 to 1600. Pre-filtering does not depend on $\ef$ and
appears as a horizontal line. Dotted verticals mark recall 0.85, 0.9 and 0.95.
The recall axis starts at 0.6, so a method whose recall stays below it is
missing from that row.}
\label{fig:sweep}
\end{figure*}

\begin{figure}[t]
\centering
\IfFileExists{6-eval/figs/sec61_sel.pdf}{\includegraphics[width=\columnwidth]{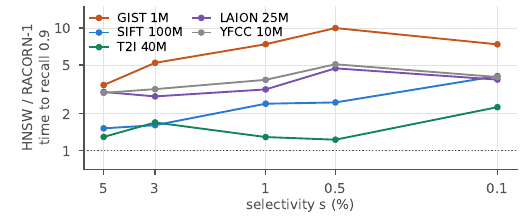}}{\fbox{\parbox{\dimexpr\columnwidth-2\fboxsep-2\fboxrule}{\centering\vspace{0.6in}figure pending\vspace{0.6in}}}}
\caption{Time to recall 0.9, HNSW over \racornone{}, against selectivity on
the five datasets of Fig.~\ref{fig:sweep}; a ratio above the dotted line
means \racornone{} is the quicker. Both methods reach 0.9 on every cell
shown.}
\label{fig:sel}
\end{figure}

The sweep reports the five datasets at their larger index, at selectivities
from 5\% to 0.1\%. Fig.~\ref{fig:sweep} shows the four rows from 3\% down.
As selectivity falls, \acornone{} and NaviX lose recall. From 1\% down they
do not reach 0.9 at any $\ef$, and at 0.1\% their recall stays below 0.03. HNSW keeps its recall, but its latency grows
with $1/\sel$. \racornone{} reaches a recall above 0.95 on every cell of the
sweep. From 3\% down it has the best QPS--recall curve of the
predicate-agnostic family. We read the twenty cells of Fig.~\ref{fig:sweep} at
three recall targets.

At recall 0.9, \racornone{} is faster than HNSW on all twenty cells, by
1.2--10.0$\times$ (Fig.~\ref{fig:sel}). At 1\% the lead is \mbox{1.3--8.4}$\times$ on the
eight indexes of Table~\ref{tab:datasets}. The lead widens on every cell as the recall target falls to 0.85, because
HNSW has no setting that stops at 0.85. Even at $\ef = 100$, the lowest $\ef$
of the sweep, it returns more than 0.85 on every cell. For a latency-sensitive
service that accepts 0.85, the time HNSW spends above 0.85 is therefore an
added cost. The lead shrinks as the target rises to 0.95, because HNSW expands failing and
passing nodes alike and therefore has the higher recall at equal $\ef$
(\S\ref{sec:cost}). To reach a higher target \racornone{} raises $\ef$
further than HNSW does, and pays for it in cycles. On three cells HNSW
reaches 0.95 first.
The ratio depends more on the dataset than on selectivity, because the costs
of both methods grow with $1/\sel$. In Fig.~\ref{fig:sel}, GIST~1M has the
highest ratio at every selectivity and Text2Image~40M is near the bottom.

\begin{figure*}[t]
\centering
\figorpending{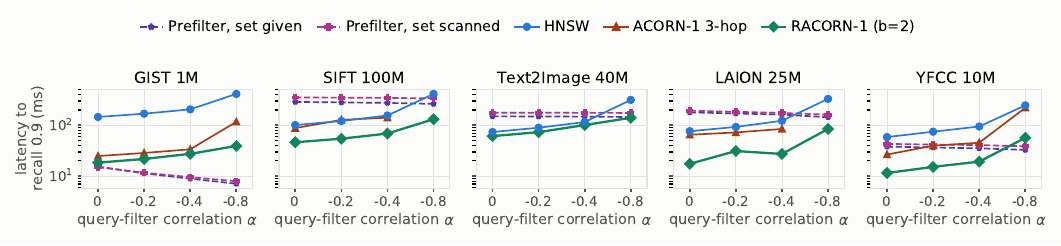}{0.5in}
\caption{Latency at which each method first reaches recall 0.9 at 1\%
selectivity, against the correlation $\alpha$ between predicate and query,
from uncorrelated ($\alpha = 0$) to $\alpha = -0.8$, where four fifths of the
passing keys are the farthest from the query. A method with no point at
an $\alpha$ does not reach 0.9 there at any $\ef$: \acornone{} and NaviX have
none, and the three-hop variant has none on Text2Image~40M. Pre-filtering does
not depend on $\alpha$.}
\label{fig:corr}
\end{figure*}

Pre-filtering with the set given is the fastest method at 0.1\% on every
dataset, 6--238$\times$ faster than \racornone{} at recall 0.9. The margin
narrows from 1M to 100M vectors, because pre-filtering computes a distance to
each of the $N\sel$ passing vectors. The passing set is not free, however.
The scanned form obtains it by evaluating the predicate on all $N$ keys,
regardless of selectivity. At 0.1\% this form is faster than \racornone{} on
GIST~1M but slower on SIFT~100M. The crossover depends on the dimension and
the graph as well as on $N$. On our datasets it lies between YFCC~10M and
LAION~25M. \S\ref{sec:eval-inexpressible} measures predicates whose
evaluation is itself expensive.

The three-hop variant recovers only part of what \acornone{} loses. It does
not reach 0.9 on any dataset at 0.1\%, nor on Text2Image~40M below 3\%. On
the other cells it reaches 0.9 but takes 1.3--4.6$\times$ longer than
\racornone{}. Each short expansion examines most of its three-hop
neighborhood, and the next one examines many of the same nodes again
(\S\ref{sec:asf}). NaviX's adaptive-local heuristic
reaches higher recall than \acornone{} from 1\% down. Like \acornone{}, it
adds only passing nodes to the candidate queue, so it does not reach 0.9
from 1\% down and collapses at 0.1\%. From 3\% down the choice is between HNSW and
\racornone{}, and \racornone{} reaches 0.85 and 0.9 first on every cell.

%% file: 6-eval/62-correlation.tex
Correlation between predicate and query is the second axis of filtered
search after selectivity~\cite{acorn,fvs-survey}. ACORN compares the distance
from the query to the passing set and to a random set of the same size. VecBench~\cite{vecbench} and NaviX~\cite{navix} base their measures
on the fraction of the query's $k$ nearest neighbors that pass. We use a rank
statistic over the whole passing set instead, which does not depend on $k$,
\begin{equation}
  c_r = 1 - \frac{2}{|A|}\sum_{x \in A} \frac{r(x)}{N},
  \label{eq:corr}
\end{equation}
where $A$ is the passing set and $r(x)$ counts the base vectors closer to the
query than $x$. $c_r$ is 0 for an uncorrelated set and $-1$ when every
passing key is the farthest. \S\ref{sec:eval-natural} places the natural filters on
the same scale.

The predicate of this section builds a passing set of $N\sel$ keys for each
query. A fraction $|\alpha|$ of the keys comes from the far end of the
query's distance ranking and the rest is drawn uniformly, so selectivity
and correlation can be set independently. The measured $c_r$ matches $\alpha$
within 0.01 on every cell. Positive correlation puts passing
keys near the query and is the easy direction for every method, so we
report the adverse one. The sweep holds selectivity at 1\% and 0.1\% and
moves $\alpha$ from 0 to $-0.8$. Fig.~\ref{fig:corr} shows the 1\% cells.

Moving passing keys away from the query slows HNSW by a factor set by $\alpha$.
Only the uniformly drawn part of the passing set, a fraction $1-|\alpha|$
of it, can lie near the query, so HNSW's cycle count rises by a factor of
$1/(1-|\alpha|)$ over its uncorrelated count $\ef/\sel$, $5\times$ at
$\alpha = -0.8$. The measured ratios lie within 0.07 of this factor on every
dataset at both selectivities. This is the dependence on correlation that
\S\ref{sec:cost} set aside. Correlation alone makes \acornone{} and NaviX
collapse: at 1\% their highest recall falls below 0.05 at $\alpha = -0.8$.
At that $\alpha$, \racornone{} reaches 0.9 on every cell, 2.2--10.5$\times$
sooner than HNSW. Adverse correlation slows HNSW and \racornone{} by similar
factors (Fig.~\ref{fig:corr}).

The three-hop variant does not collapse at 1\%, but it does not always
reach 0.9. At $\alpha = -0.8$ it reaches 0.9 on half the datasets at 1\%
and on none at 0.1\%. Pre-filtering does not depend on $\alpha$, because
the given set costs $N\sel$ distance computations regardless of
correlation. At 1\% it is slower than \racornone{} at every $\alpha$ on the
three datasets of 25M vectors and more. At $\alpha = 0$ the passing set is
uncorrelated, as in \S\ref{sec:eval-sweep}, and the results agree with that
section. Adverse correlation brings the collapse of \acornone{} and NaviX from
0.1\% up to 1\%, while \racornone{} keeps its lead over HNSW.

%% file: 6-eval/63-natural.tex
\begin{figure}[t]
\centering
\IfFileExists{6-eval/figs/sec63_natural_25m.pdf}{\includegraphics[width=\columnwidth]{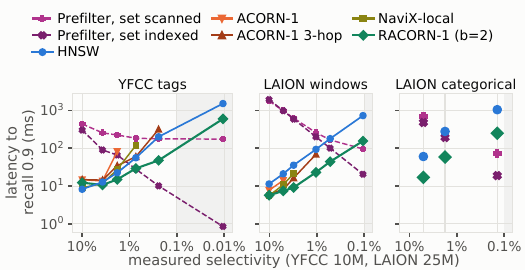}}{\fbox{\parbox{\dimexpr\columnwidth-2\fboxsep-2\fboxrule}{\centering\vspace{0.7in}figure pending\vspace{0.7in}}}}
\caption{Latency at which each method first reaches recall 0.9 on the filters that
YFCC~10M and LAION~25M ship, against the selectivity each group turns out to
have. The categorical panel holds the three equality predicates on LAION's
columns at the selectivities the data gives them, NSFW = UNSURE, NSFW = NSFW
and LICENSE = CC from left to right; the shaded region is below 0.1\%, outside
the range this paper targets. A method with no point at a group does not reach
0.9 there at any $\ef$. Pre-filtering appears with the passing set read from an
index over the tag or column and with the set scanned.}
\label{fig:natural}
\end{figure}

Two workloads replace the controlled predicate with filters the datasets
ship. YFCC~10M comes with the tag queries of the Big ANN filtered
track~\cite{bigannfiltered}, in which a query names one or two tags and
selects the vectors that carry all of them. We group its 992 queries into
six buckets by the selectivity of their tags. Over the full query set 70\% of
the queries select less than 1\%, and 47\% select less than 0.1\%, below the
range this paper targets. On LAION at 1M and 25M we build two kinds of predicate
from the metadata columns that come with the embeddings. Three are equality
predicates on its categorical columns, at the selectivities the data gives
them (Fig.~\ref{fig:natural}), and six are windows on the original pixel
count, placed at random per query and sized to selectivities from 10\% to
0.1\%.

Fig.~\ref{fig:natural} shows the time to reach recall 0.9 on the YFCC~10M
and LAION~25M groups, against their measured selectivity. \acornone{}
reaches recall 0.9 on a third of the groups and on none below 1\%, NaviX
on half of them, and the three-hop variant on two thirds. \racornone{}
reaches it on every group, and below 1\% it gets there 1.9--10.3$\times$ sooner than HNSW.

The correlation $c_r$ of (\ref{eq:corr}) puts the groups between $-0.27$
and $+0.28$. At equal selectivity it separates the two LAION predicate
types. The categorical groups have $c_r$ between $-0.17$ and $-0.27$, and
the windows have a $c_r$ of 0. At 25M \racornone{} takes 2.3$\times$ as long on the
5.1\% categorical group as on the 5.1\% window, while \acornone{} does not
reach 0.9 on the categorical group. \S\ref{sec:eval-corr} measures this
dependence on correlation.

Predicate-aware indexes are built for filters like these and outperform the
predicate-agnostic family on them, \racorn{} included (\S\ref{sec:landscape}),
so the comparison here stays within the family. Pre-filtering from an index over the tag or
column is faster than \racornone{} on every group below 0.5\%. Within the
family, the filters that the
datasets ship confirm \S\ref{sec:eval-sweep} and \S\ref{sec:eval-corr}.
\racornone{} is the only filter-first method that reaches 0.9 on every
group, and a correlated categorical group costs it more than a window of
equal selectivity.

%% file: 6-eval/64-inexpressible.tex
\begin{figure}[t]
\centering
\IfFileExists{6-eval/figs/sec64_conj_caption_hetero.pdf}{\includegraphics[width=\columnwidth]{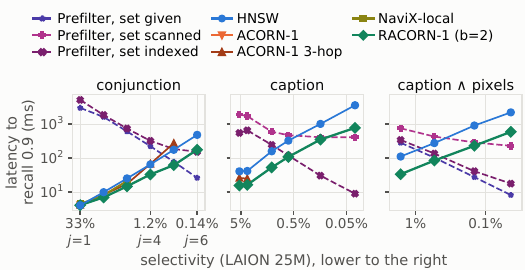}}{\fbox{\parbox{\dimexpr\columnwidth-2\fboxsep-2\fboxrule}{\centering\vspace{0.7in}figure pending\vspace{0.7in}}}}
\caption{Latency at which each method first reaches recall 0.9 on LAION~25M
under three predicates: a conjunction of the first $j$ of six synthetic
attributes, a caption keyword in six selectivity buckets, and the keyword
conjoined with a pixel-count window, in four groups. A method with no point
does not reach 0.9 there at any $\ef$. The scanned set on the conjunction and
the given set on the captions are within 14\% and 4\% of the indexed set and
not drawn. The third panel draws all three forms.}
\label{fig:conj}
\end{figure}

Four workloads put the predicate beyond a single range or label. No single
predicate-aware index covers two of them, the per-query allowlist and the caption
keyword conjoined with a pixel-count window. The allowlist is an arbitrary set
with no unit to index, and we reuse the $\alpha = 0$ runs of
\S\ref{sec:eval-corr} for it. On five datasets we run a conjunction over six
uniform attributes of cardinality three, and each query fixes the first $j$
of them. Each attribute alone stays at 33.3\%, while the joint selectivity
falls from 33.3\% to 0.137\% as $j$ runs from 1 to 6. The LAION captions give
token predicates in six selectivity buckets at 1M and 25M. The fourth
workload conjoins a caption keyword with a window on the original pixel
count, a label and a range predicate at once, in four groups from
1.64\% to 0.043\%.

\racornone{} reaches recall 0.9 on all 46 groups we run here, 30 on the
conjunction, 12 on the captions and 4 heterogeneous. \acornone{} misses 29
of them, NaviX 28 and the three-hop variant 19, and none of them reaches
0.9 on any heterogeneous group.
\acornone{} loses recall at a higher selectivity than under the uniform
predicate. On the captions its highest recall is 0.82 in the 5\% bucket
(Fig.~\ref{fig:conj}), against 0.97 at the same selectivity under the
uniform predicate of \S\ref{sec:eval-sweep} on the same index. Against HNSW,
\racornone{} reaches 0.9 sooner on every caption bucket and heterogeneous
group, by 2.5--8.7$\times$. On the conjunction it reaches 0.9 sooner on 14 of
the 15 cells at joint selectivity 1.23\% and below, by 1.3--7.2$\times$. The
exception is Text2Image~40M at 1.23\%, where HNSW is quicker by
1.21$\times$.

When the set is given, pre-filtering is again the fastest method at the
bottom of each workload, as \S\ref{sec:landscape} states: below 0.16\% on
the captions, below 0.15\% on the heterogeneous groups and at 0.1\% on the
allowlist. Only the allowlist arrives as a set, from the application that
computed it, and it has no attribute list to intersect. For the other three,
obtaining the set is part of the query.
The caption predicate is the expensive one that \S\ref{sec:cost} set aside:
scanning it is 14--47$\times$ slower than the given set at 25M below 0.5\%
selectivity, while a posting index over the tokens brings it back to the
given set. Attribute lists do not make the conjunction cheap, because each
attribute alone selects a third of the keys and intersecting six lists does
the work of a scan. The intersection grows with $N$, unlike a graph query,
so at 0.137\% pre-filtering from the intersected lists takes nearly as long
as a \racornone{} query on LAION~25M and 3.2$\times$ as long on SIFT~100M.

Fig.~\ref{fig:conj} shows the cost of choosing a method for the conjunction
from the per-attribute selectivities. On LAION~25M, from $j = 1$ to $j = 6$ the time to recall 0.9
falls 114$\times$ for the given set and rises 118$\times$ for HNSW and
43$\times$ for \racornone{}, which holds 0.9 at every $j$. Beyond a single attribute, the comparison
thus stays inside the family. There \racornone{} is the only filter-first
method that holds recall 0.9 on every group, and from 3\% down it reaches
0.9 sooner than HNSW on every group but one.

%% file: 6-eval/65-det.tex
\begin{figure}[t]
\centering
\IfFileExists{6-eval/figs/sec65_det.pdf}{\includegraphics[width=\columnwidth]{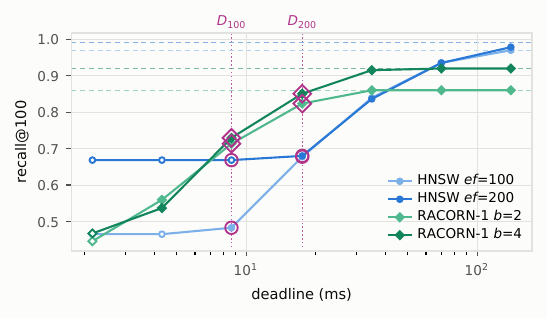}}{\fbox{\parbox{\dimexpr\columnwidth-2\fboxsep-2\fboxrule}{\centering\vspace{0.7in}figure pending\vspace{0.7in}}}}
\caption{Recall against the deadline on SIFT~100M at 1\% selectivity, HNSW
at $\ef$ 100 and 200 and \racornone{} at $\ef = 200$ with $\br = 2$ and $\br = 4$, all
with DET. Ringed points mark $D_{100}$ and $D_{200}$, the deadlines at which HNSW
at $\ef$ 100 and 200 fills $W$, and dashed lines are each series' recall
without a deadline, which a series meets once it finishes before its deadline.
Open markers are deadlines that the series could not hold, because its
measured latency exceeded them.}
\label{fig:det}
\end{figure}

We give HNSW and \racornone{} the same deadline and compare the recall they
return. The deadlines are set relative to HNSW. $D_{100}$ and $D_{200}$ are
the median times at which HNSW at $\ef = 100$ and $200$ fills $W$, measured
without a deadline in the same process. The seven deadlines run from
$D_{100}/4$ to $8D_{200}$. The grid is the eight datasets at 3\%, 1\%,
0.5\% and 0.1\%, with HNSW at the same two $\ef$ and \racornone{} at
$\ef = 200$ with $\br$ from 1 to 8. At each deadline a method takes the
setting that returns the highest recall within the deadline.

Fig.~\ref{fig:det} shows one cell. At $D_{100}$ HNSW has just filled $W$, so
its recall is still that of its first $\ef$ results (\S\ref{sec:cost}). By
then \racornone{} has been converging for most of the deadline. Past
$D_{200}$ HNSW runs its convergence phase, and its curve climbs to meet
\racornone{}'s and then passes it. \racornone{}'s curve is already flat
there, because it has converged.

At $D_{100}$ \racornone{} returns the higher recall on every cell, by a
median of 0.28 and at least 0.18. At $D_{200}$ it returns the higher
recall on every cell as well. At both deadlines the median ratio of their measured
latencies is 1.00, so the comparison is at equal time and not only at
equal deadline. Under a deadline a larger $\br$ is not always better,
because the extra distance computations it adds (\S\ref{sec:asf}) need time
to raise the recall. The best $\br$ for a cell grows with the deadline. It
is 1 at $D_{100}/4$ on every cell, and on nearly every cell it is 2 or 4 at
$D_{100}$, 4 to 8 at $D_{200}$, then 8 from $2D_{200}$ up. The fixed $\br = 2$ of the
main experiments comes within 0.03 of the best single $\br$ at both deadlines. Given the same time, \racornone{} returns the higher
recall on every cell at both deadlines, and HNSW recovers the difference
only with more time.

%% file: 6-eval/66-ablation.tex
\begin{figure}[t]
\centering
\IfFileExists{6-eval/figs/sec66_ablation.pdf}{\includegraphics[width=\columnwidth]{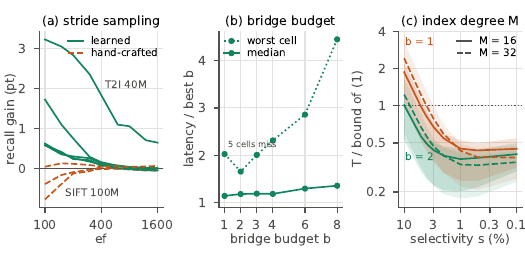}}{\fbox{\parbox{\dimexpr\columnwidth-2\fboxsep-2\fboxrule}{\centering\vspace{0.7in}figure pending\vspace{0.7in}}}}
\caption{The two parameters and the index degree, on the eight indexes.
(a) The recall stride sampling adds over prefix truncation at 1\%
selectivity and $\br = 2$, against $\ef$, solid for the five datasets with
learned embeddings and dashed for the three with hand-crafted descriptors. (b) The latency
of one fixed $\br$ against the best of $\br \in \{1, 2, 3, 4, 6, 8\}$ for each of the eight datasets at
3\% selectivity and below and at recall targets 0.9 and 0.95: the median cell
and the worst. At $\br = 1$ five cells do not reach recall 0.95 at any $\ef$.
(c) $T/$bound of~(\ref{eq:bound}) against selectivity at $M = 16$ (solid)
and $32$ (dashed), $\br = 1$ and $2$: median and range over the eight
datasets.}
\label{fig:ablation}
\end{figure}

The value of stride sampling depends on the embedding. At 1\%, 0.5\% and
0.1\% selectivity with $\ef = 100$, the five datasets with learned embeddings gain up
to 0.033 recall over prefix truncation, and the three with hand-crafted
descriptors lose up to 0.008 (Fig.~\ref{fig:ablation}a). The split does not come from the
distance function: GIST~1M, whose vector norms vary sixfold, stays on the
hand-crafted side when searched with cosine instead of $L_2$. The gain shrinks as $\ef$ grows.

We leave stride sampling on by default, because the learned embeddings gain
more than the hand-crafted descriptors lose. The figures average over
predicate seeds, and on every learned cell the gain keeps its sign on five
independently built graphs. On \acornone{} the sampler acts only at the
two-hop cap, which these selectivities rarely reach (\S\ref{sec:stride}), so
its recall with and without stride is identical in all measurements we
made.

Fig.~\ref{fig:ablation}b shows why the main experiments fix $\br = 2$. Over the
eight datasets at four selectivities and two recall targets, a
fixed $\br = 2$ reaches every recall target and is at most 1.66$\times$
slower than the best $\br$ for each cell. That is the smallest worst case
of the six values. $\br = 1$ misses recall 0.95 on five cells,
and $\br \ge 4$ is up to 4.45$\times$ slower
in its worst cell.

At $M = 16$ the bound~(\ref{eq:bound}) holds from 3\% down at every $\br$ of the
sweep. Fig.~\ref{fig:ablation}c
repeats the measurement on the eight indexes rebuilt at $M = 32$, with
$\ef = 100$. Their $\Mavg$ is 1.2--1.5$\times$ higher, so each expansion
tests more two-hop nodes, and the $T/$bound curves shift toward lower
selectivity. The bound holds from 5\% down at $\br = 2$ but only from 2\%
down at $\br = 1$. The defaults of the main experiments, stride on and $\br = 2$,
thus have the smallest worst case, and the 3\% boundary moves with $M$.

%% file: 8-conclusion/conclusion.tex
Below a few percent selectivity, filtered search on an HNSW graph built
without metadata breaks down. \acornone{} runs out of passing candidates
and collapses below 1\%. HNSW keeps its recall but spends about
$\ef/\sel$ expansion cycles. \racornone{} keeps the graph and the
filter-first traversal, and its Adaptive Search Fallback fills each
expansion up to a target with failing nodes, under a per-expansion budget.
The search expands these nodes as bridges but never returns them. With
bridges the traversal no longer starves: the recall of \racornone{} reaches
0.95 at every selectivity of the sweep, while that of \acornone{} stays below
0.9 at every $\ef$ from 1\% down.

Unlike \acornone{}, whose candidate queue runs dry, \racornone{} fills the
result queue, and at
the same $\ef$ as HNSW it has lower recall but several times lower latency, a middle ground between
\acornone{} and HNSW.
At 1\% selectivity and a recall target of 0.9, \racornone{} is 1.3--8.4$\times$ faster
than HNSW on the eight indexes, and it remains faster under adverse
correlation. It reaches 0.9 on every natural filter and every predicate
beyond a single attribute we evaluated, and it gets there sooner than HNSW.
We also ran both methods under
deadline-based early termination with the same deadline, the time HNSW at
$\ef = 100$ needs to fill its result queue. At that deadline \racornone{}
ends with the higher recall on every dataset and selectivity, at measured
latencies equal to HNSW's. Predicate-aware indexes remain faster on the range and label
predicates they are built for. For predicates that no index was built for, \racorn{} extends the \acornone{}
a system already runs without changing its index, which holds no metadata to
update. From 3\% down \racornone{} has the best QPS--recall curve of the
predicate-agnostic family.

%% file: 9-ack/ack.tex
Claude (Anthropic), used through Claude Code, drafted the text of this
paper from the authors' outlines, measurements and notes, and wrote and
ran the experiment, analysis and plotting scripts under the authors'
direction. The authors designed the study and the
experiments, wrote the \acornone{} and \racornone{} implementations, and
reviewed and revised every sentence.

%% file: main.bbl
\begin{thebibliography}{10}
\providecommand{\url}[1]{#1}
\csname url@samestyle\endcsname
\providecommand{\newblock}{\relax}
\providecommand{\bibinfo}[2]{#2}
\providecommand{\BIBentrySTDinterwordspacing}{\spaceskip=0pt\relax}
\providecommand{\BIBentryALTinterwordstretchfactor}{4}
\providecommand{\BIBentryALTinterwordspacing}{\spaceskip=\fontdimen2\font plus
\BIBentryALTinterwordstretchfactor\fontdimen3\font minus
  \fontdimen4\font\relax}
\providecommand{\BIBforeignlanguage}[2]{{%
\expandafter\ifx\csname l@#1\endcsname\relax
\typeout{** WARNING: IEEEtran.bst: No hyphenation pattern has been}%
\typeout{** loaded for the language `#1'. Using the pattern for}%
\typeout{** the default language instead.}%
\else
\language=\csname l@#1\endcsname
\fi
#2}}
\providecommand{\BIBdecl}{\relax}
\BIBdecl

\bibitem{rag}
P.~Lewis \emph{et~al.}, ``Retrieval-augmented generation for
  knowledge-intensive {NLP} tasks,'' in \emph{Advances in Neural Information
  Processing Systems (NeurIPS)}, vol.~33, 2020, pp. 9459--9474.

\bibitem{fvs-survey}
Y.~Chronis, H.~Caminal, Y.~Papakonstantinou, F.~\"{O}zcan, and A.~Ailamaki,
  ``Filtered vector search: State-of-the-art and research opportunities,''
  \emph{Proc. VLDB Endow.}, vol.~18, no.~12, pp. 5488--5492, 2025.

\bibitem{acorn}
L.~Patel, P.~Kraft, C.~Guestrin, and M.~Zaharia, ``{ACORN}: Performant and
  predicate-agnostic search over vector embeddings and structured data,''
  \emph{Proc. ACM Manag. Data}, vol.~2, no.~3, pp. 120:1--120:27, 2024.

\bibitem{hnsw}
Y.~A. Malkov and D.~A. Yashunin, ``Efficient and robust approximate nearest
  neighbor search using hierarchical navigable small world graphs,'' \emph{IEEE
  Trans. Pattern Anal. Mach. Intell.}, vol.~42, no.~4, pp. 824--836, 2020.

\bibitem{lucene}
B.~Trent, ``Filtered {HNSW} search, fast mode,''
  \url{https://www.elastic.co/search-labs/blog/filtered-hnsw-knn-search},
  {Elastic Search Labs}, 27 February 2025; describes the change shipped in
  {Lucene} 10.2. Accessed 2026-10-04.

\bibitem{weaviate}
{Weaviate}, ``Filtering,''
  \url{https://docs.weaviate.io/weaviate/concepts/filtering}, {Weaviate}
  documentation. Accessed 2026-10-04.

\bibitem{qdrant}
{Qdrant}, ``{Similarity Search},''
  \url{https://qdrant.tech/documentation/search/search}, {Qdrant}
  documentation. Accessed 2026-10-04.

\bibitem{vespablog}
J.~B{\"o}ker, ``Additions to {HNSW} in {Vespa}: {ACORN-1} and adaptive beam
  search,'' \url{https://blog.vespa.ai/additions-to-hnsw/}, blog post, 4
  September 2025. Accessed 2026-10-04.

\bibitem{bigannfiltered}
H.~V. Simhadri \emph{et~al.}, ``Results of the {Big ANN}: {NeurIPS'23}
  competition,'' in \emph{Advances in Neural Information Processing Systems
  (NeurIPS), Datasets and Benchmarks Track}, vol.~38, 2025.

\bibitem{laion400m}
C.~Schuhmann \emph{et~al.}, ``{LAION-400M}: Open dataset of {CLIP}-filtered 400
  million image-text pairs,'' \emph{CoRR}, vol. abs/2111.02114, 2021.

\bibitem{serf}
C.~Zuo, M.~Qiao, W.~Zhou, F.~Li, and D.~Deng, ``{SeRF}: Segment graph for
  range-filtering approximate nearest neighbor search,'' \emph{Proc. ACM Manag.
  Data}, vol.~2, no.~1, pp. 69:1--69:26, 2024.

\bibitem{irangegraph}
Y.~Xu, J.~Gao, Y.~Gou, C.~Long, and C.~S. Jensen, ``{iRangeGraph}: Improvising
  range-dedicated graphs for range-filtering nearest neighbor search,''
  \emph{Proc. ACM Manag. Data}, vol.~2, no.~6, pp. 239:1--239:26, 2024.

\bibitem{filtereddiskann}
S.~Gollapudi \emph{et~al.}, ``{Filtered-DiskANN}: Graph algorithms for
  approximate nearest neighbor search with filters,'' in \emph{Proc. ACM Web
  Conference (WWW)}, 2023, pp. 3406--3416.

\bibitem{ung}
Y.~Cai, J.~Shi, Y.~Chen, and W.~Zheng, ``Navigating labels and vectors: A
  unified approach to filtered approximate nearest neighbor search,''
  \emph{Proc. ACM Manag. Data}, vol.~2, no.~6, pp. 246:1--246:27, 2024.

\bibitem{navix}
G.~Sehgal and S.~Saliho{\u{g}}lu, ``{NaviX}: A native vector index design for
  graph {DBMSs} with robust predicate-agnostic search performance,''
  \emph{Proc. VLDB Endow.}, vol.~18, no.~11, pp. 4438--4450, 2025.

\bibitem{diskann}
S.~J. Subramanya, Devvrit, H.~V. Simhadri, R.~Krishnaswamy, and R.~Kadekodi,
  ``{DiskANN}: Fast accurate billion-point nearest neighbor search on a single
  node,'' in \emph{Advances in Neural Information Processing Systems
  (NeurIPS)}, vol.~32, 2019.

\bibitem{nsg}
C.~Fu, C.~Xiang, C.~Wang, and D.~Cai, ``Fast approximate nearest neighbor
  search with the navigating spreading-out graph,'' \emph{Proc. VLDB Endow.},
  vol.~12, no.~5, pp. 461--474, 2019.

\bibitem{faiss}
M.~Douze \emph{et~al.}, ``The {Faiss} library,'' \emph{CoRR}, vol.
  abs/2401.08281, 2024.

\bibitem{hnswlib}
Y.~A. Malkov \emph{et~al.}, ``hnswlib: Header-only {C++}/{Python} library for
  fast approximate nearest neighbors,''
  \url{https://github.com/nmslib/hnswlib}, accessed 2026-10-04.

\bibitem{usearch}
A.~Vardanian, ``{USearch}: Smaller \& faster similarity search \& clustering
  engine,'' \url{https://github.com/unum-cloud/usearch}, 2023, accessed
  2026-10-04.

\bibitem{sift}
H.~J{\'e}gou, M.~Douze, and C.~Schmid, ``Product quantization for nearest
  neighbor search,'' \emph{IEEE Trans. Pattern Anal. Mach. Intell.}, vol.~33,
  no.~1, pp. 117--128, 2011.

\bibitem{bigann1b}
H.~J{\'e}gou, R.~Tavenard, M.~Douze, and L.~Amsaleg, ``Searching in one billion
  vectors: Re-rank with source coding,'' in \emph{IEEE International Conference
  on Acoustics, Speech and Signal Processing (ICASSP)}, 2011, pp. 861--864.

\bibitem{yfcc}
B.~Thomee \emph{et~al.}, ``{YFCC100M}: The new data in multimedia research,''
  \emph{Commun. ACM}, vol.~59, no.~2, pp. 64--73, 2016.

\bibitem{bigann}
H.~V. Simhadri \emph{et~al.}, ``Results of the {NeurIPS'21} challenge on
  billion-scale approximate nearest neighbor search,'' in \emph{Proceedings of
  the NeurIPS 2021 Competitions and Demonstrations Track}, ser. Proceedings of
  Machine Learning Research, vol. 176, 2022, pp. 177--189.

\bibitem{vbase}
Q.~Zhang \emph{et~al.}, ``{VBASE}: Unifying online vector similarity search and
  relational queries via relaxed monotonicity,'' in \emph{Proc. USENIX
  Symposium on Operating Systems Design and Implementation (OSDI)}, 2023, pp.
  377--395.

\bibitem{digra}
M.~Jiang \emph{et~al.}, ``{DIGRA}: A dynamic graph indexing for approximate
  nearest neighbor search with range filter,'' \emph{Proc. ACM Manag. Data},
  vol.~3, no.~3, pp. 148:1--148:26, 2025.

\bibitem{vecflow}
J.~Xi, C.~Mo, B.~Karsin, A.~Chirkin, M.~Li, and M.~Zhang, ``{VecFlow}: A
  high-performance vector data management system for filtered-search on
  {GPUs},'' \emph{Proc. ACM Manag. Data}, vol.~3, no.~4, pp. 271:1--271:27,
  2025.

\bibitem{learned-et}
C.~Li, M.~Zhang, D.~G. Andersen, and Y.~He, ``Improving approximate nearest
  neighbor search through learned adaptive early termination,'' in \emph{Proc.
  ACM SIGMOD International Conference on Management of Data}, 2020, pp.
  2539--2554.

\bibitem{darth}
M.~Chatzakis, Y.~Papakonstantinou, and T.~Palpanas, ``{DARTH}: Declarative
  recall through early termination for approximate nearest neighbor search,''
  \emph{Proc. ACM Manag. Data}, vol.~3, no.~4, pp. 242:1--242:26, 2025.

\bibitem{clip}
A.~Radford \emph{et~al.}, ``Learning transferable visual models from natural
  language supervision,'' in \emph{Proc. International Conference on Machine
  Learning (ICML)}, ser. Proceedings of Machine Learning Research, vol. 139,
  2021, pp. 8748--8763.

\bibitem{lucene-hnsw}
{Apache Lucene}, ``Class {HnswGraphBuilder},''
  \url{https://lucene.apache.org/core/10_5_2/core/org/apache/lucene/util/hnsw/HnswGraphBuilder.html},
  {Lucene} 10.5.2 API documentation; \texttt{DEFAULT\_MAX\_CONN} = 16,
  \texttt{DEFAULT\_BEAM\_WIDTH} = 100. Accessed 2026-10-18.

\bibitem{qdrant-indexing}
{Qdrant}, ``Indexing,''
  \url{https://qdrant.tech/documentation/manage-data/indexing/}, {Qdrant}
  documentation; \texttt{m} = 16, \texttt{ef\_construct} = 100 by default.
  Accessed 2026-11-01.

\bibitem{navix-artifact}
G.~Sehgal, ``{Kuzu Navix},'' \url{https://github.com/gaurav8297/kuzu}, 2025,
  artifact of the NaviX paper, commit 5f149a2. Accessed 2026-10-20.

\bibitem{numkong}
A.~Vardanian, ``{NumKong}: 2000 mixed precision kernels for all,''
  \url{https://github.com/ashvardanian/NumKong}, 2026, version 7.2.0. Accessed
  2026-10-20.

\bibitem{vecbench}
X.~Zhang, C.~Zhang, J.~Fan, G.~Li, and X.~Du, ``{VecBench}: A controllable
  benchmark for filtered vector search,'' \emph{Proc. ACM Manag. Data}, vol.~4,
  no.~3, pp. 248:1--248:27, 2026.

\end{thebibliography}
